# On-instrument wavefront sensor design for the TMT infrared imaging spectrograph (IRIS) update

Jennifer Dunn*[a], Vlad Reshetov[a], Jenny Atwood[a], John Pazder[a], Bob Wooff[a],
David Loop[a], Leslie Saddlemyer[a],
Anna M. Moore[b], James E. Larkin[c],
[a]National Research Council Herzberg, 5071 W. Saanich Rd., Victoria, BC, V9E 2E7, Canada
[b]Caltech Optical Observatories,1200 E California Blvd., Mail Code 11-17, Pasadena, CA 91125;
[c]Department of Physics and Astronomy, University of California, Los Angeles, CA 90095-1547;

## ABSTRACT

The first light instrument on the Thirty Meter Telescope (TMT) project will be the InfraRed Imaging Spectrograph (IRIS). IRIS will be mounted on a bottom port of the facility AO instrument NFIRAOS. IRIS will report guiding information to the NFIRAOS through the On-Instrument Wavefront Sensor (OIWFS) that is part of IRIS. This will be in a self-contained compartment of IRIS and will provide three deployable wavefront sensor probe arms. This entire unit will be rotated to provide field de-rotation. Currently in our preliminary design stage our efforts have included: prototyping of the probe arm to determine the accuracy of this critical component, handling cart design and reviewing different types of glass for the atmospheric dispersion.

## 1. INTRODUCTION

The InfraRed (0.84 – 2.4 micron) Imaging Spectrograph (IRIS) will be a first light instrument on the Thirty Meter Telescope (TMT) project. IRIS will utilize the advanced adaptive optics system (Narrow Field InfraRed Adaptive Optics System, NFIRAOS) and will also take advantage of an integrated On-Instrument Wavefront Sensor (OIWFS). The IRIS OIWFS uses a low-order Shack Hartmann wavefront sensor to make measurements and calculate estimations of tip/tilt for fast guiding, and focus. There will be three probes that will patrol a large 4.4 arcsec diameter field of view that has been AO corrected using NFIRAOS. It will also provide atmospheric dispersion correction on each of the probe arms. IRIS will be mounted below NFIRAOS and will rotate in place using a large rotary bearing to provide field de-rotation. The IRIS OIWFS will be maintained at the same temperature as NFIRAOS (-30 C).

NRC Herzberg is responsible for the IRIS OIWFS design, the rotator for the entire instrument, and the handling cart. In this paper we will provide an update on the recent design prototype of the OIWFS and changes done to the system since last reported in paper by Loop et al [1]. These include the prototyping of a single probe arm assembly, handling cart design and glass selection for the atmospheric dispersion corrector (ADC). The latest update for IRIS as a whole is reported this year in the paper by Moore et al [2].

---

* Jennifer.Dunn@nrc-cnrc.gc.ca phone 1 250 363-6912; fax 1 250 363-0045;

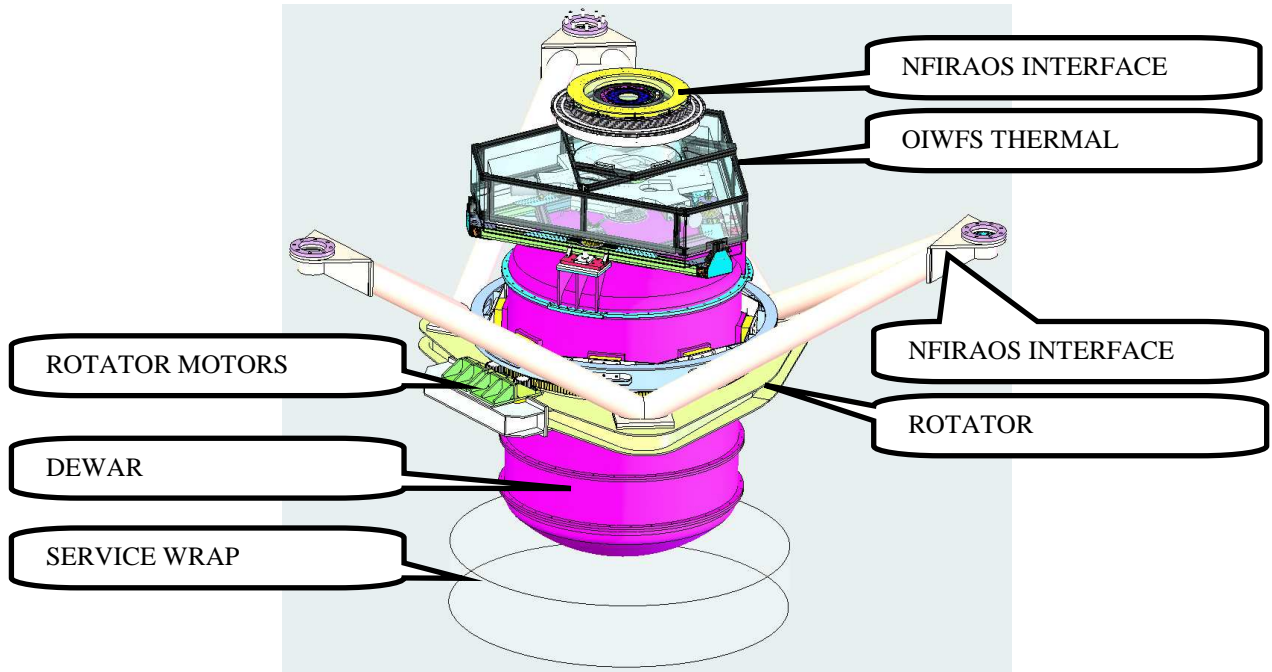

**Figure 1. IRIS Mechanical Layout**

## 2. PROBE ARM PROTOTYPE

### 2.1 OVERVIEW OF THE OIWFS PROBE ARM DESIGN

Mechanical design of the probe arms is based on a Theta-R concept: one rotational and one linear degree of freedom. As shown in Figure 2, light coming from NFIRAOS encounters the converter lens first. This element is located at the tip of the probe arm. Following the converter element, the light encounters the probe arm fold mirror, an aperture stop and a collimator lens. All four elements travel with the probe arm. The collimator lens however has an independent degree of motion: it travels 4 mm in relation to the probe arm to compensate for the field curvature of the focal plane. Further down the optical path we have a carriage with two fold mirrors (trombone mirrors). The trombone mirror carriage travels on the same linear guides as the probe arm, but at exactly half the rate of the probe arm. The position of the trombone mirrors has different offsets relative to the probe arm carriage for the imaging and for the WFS modes. Following the beam path, after the trombone mirrors, there are two static fold mirrors. The bottom fold mirror rotates with the probe arm, and the top fold mirror remains stationary.

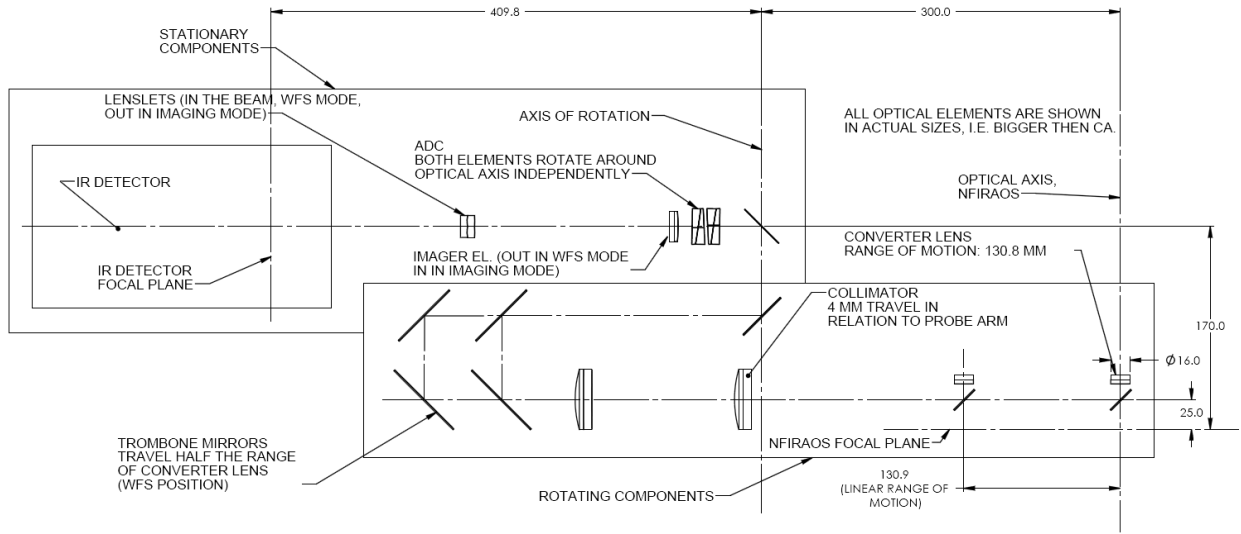

**Figure 2. Schematic layout of the probe arm assembly**

The components above the probe arm, labeled on Figure 2 as stationary, do not move with the probe arm. However, they do have their individual internal degrees of motion: two ADC elements must rotate 360° independently of each other, and the lenslet array and the imager elements must move between two fixed positions: in and out. The lenslet array element is in the "in" position in the tip-tilt-focus (TTF) mode, and "out" in the tip-tilt (TT) mode. The imager element is "in" in the TT mode, and "out" in the TTF mode. The stationary fold mirror, the ADC elements, the imaging element, and the lenslet array are combined into a stand-alone module referred to as the ADC subassembly. The intent here is to be able to assemble and integrate the ADC subassemblies outside of the OIWFS. The OIWFS IR detector is the final element in the optical path.

## 2.2 OVERVIEW OF THE PROBE ARM PROTOTYPE

The main goal of OIWFS prototype study was to verify repeatability of the positioning of the probe arm at the operating temperature of -30C. For this purpose a test stand was designed (Assembled probe arm prototype., Figure 4). The test prototype incorporates one probe arm suspended from an aluminum platform. The platform is supported from the base by three aluminum flexures. The OIWFS arm subassembly included the following actuators: probe arm linear actuator, collimator linear actuator (Figure 5), trombone mirrors linear actuator, probe arm rotary actuator.

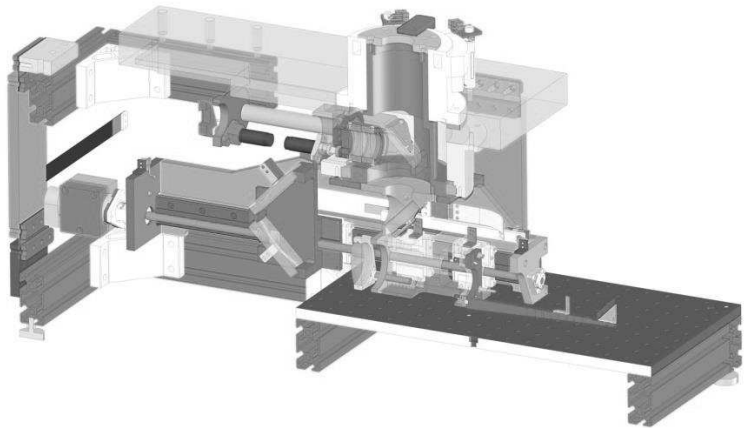

**Figure 3. Probe arm prototype, solid model, cross-section view.**

Components were assembled and verified at the room temperature first, and then tested in the cold room at -30C.

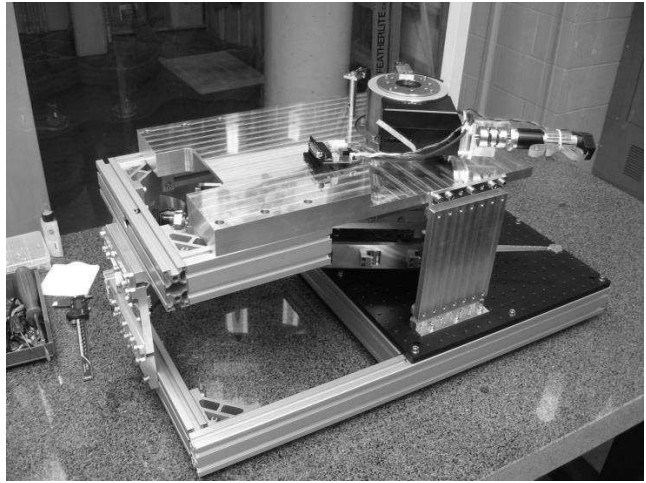

**Figure 4. Assembled probe arm prototype.**

### 2.3 TESTS OVERVIEW

The tests of the collimator assembly were done outside of the OIWFS system (Figure 5, photo on the left). The collimator assembly was clamped to an aluminum extrusion and the position of the collimator optical housing was measured using a Heidenhain linear probe. A test script sent commands to a Galil controller to move the actuator into certain positions then when the motion was completed the reading from the Heidenhain probe was logged into a file. This configuration was used for both room temperature and cold temperature environments.

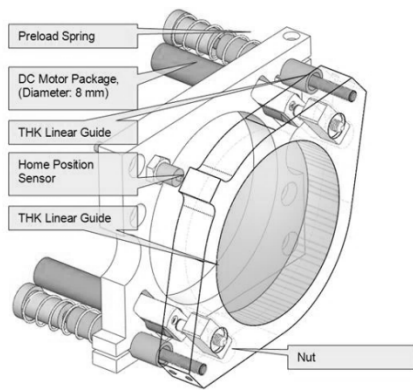 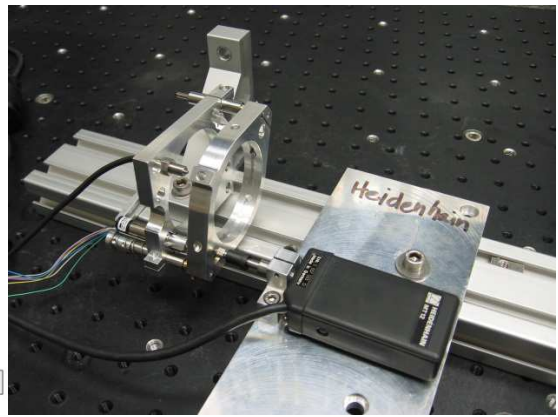

**Figure 5. Main components of the collimator assembly and photo of the prototype of the collimator assembly**

Room temperature tests of the OIWFS arm assembly were done in two stages. In the first stage the tests were performed on the Mitutoyo CMM BRTSTRATO-707. The uncertainty of probing error on this machine is 0.14 µm, and the maximum probing error is 2.16 µm. For the repeatability test only the uncertainty of probing error is of importance. The 0.14 µm is at least one order of magnitude better than the parameter we are measuring.

The OIWFS prototype arm was moved from a random position in x-y to a fixed position for measurement. The fixed position selected for measurement was located in the middle of the travel of both the rotator and the linear actuator. The cyclical script running on the CMM controller in automatic mode (thousands of cycles) measured the position of the arm by measuring the central hole in the OIWFS tip. The position of the center of the hole was averaged from six individual measurement points. Next, the CMM script gave commands to the OIWFS controller to move the arm into a random position. At this point the CMM measured a reference steel sphere, 30.4 mm in diameter, located right under the fixed measurement position of the OIWFS tip. Thus we could compensate the results for the drift of the prototype assembly relative to the CMM table. This measurement setup is shown in Figure 6.

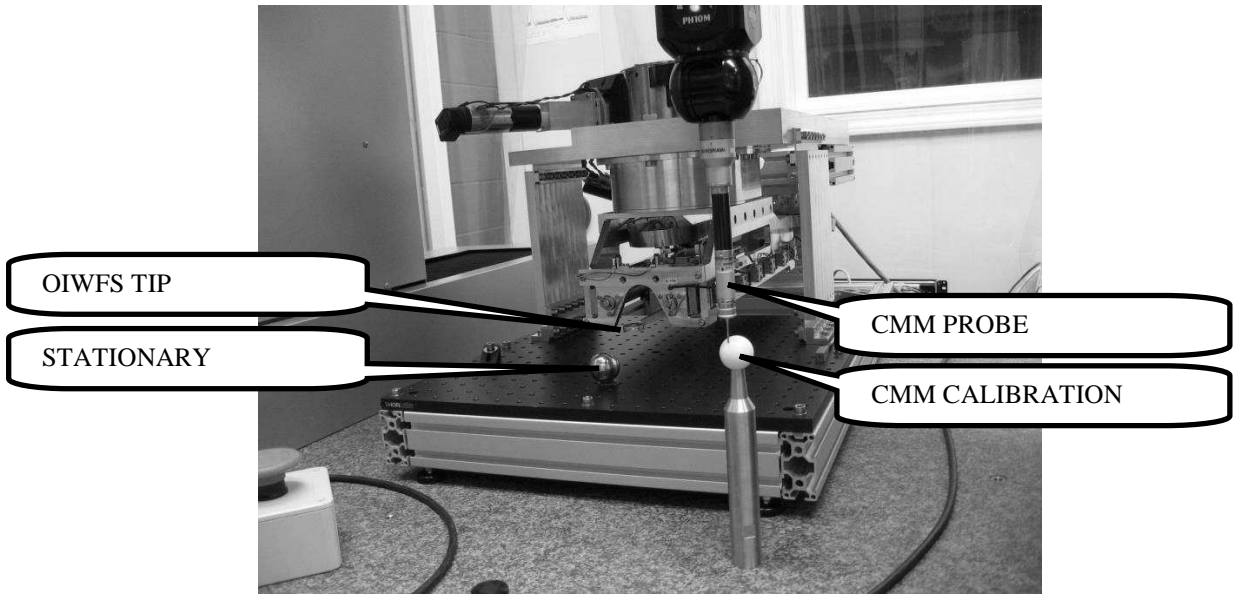

**Figure 6. Warm repeatability test using CMM**

The second type of measurement was done using a Renishaw XL-80 laser interferometer system. The XL-80 system was used to measure the repeatability of the probe arm first at room temperature, and then in the cold conditions (-32…-33 °C). A typical setup for such a measurement is schematically shown on Figure 7. In Figure 8 we show the physical layout of the measurement.

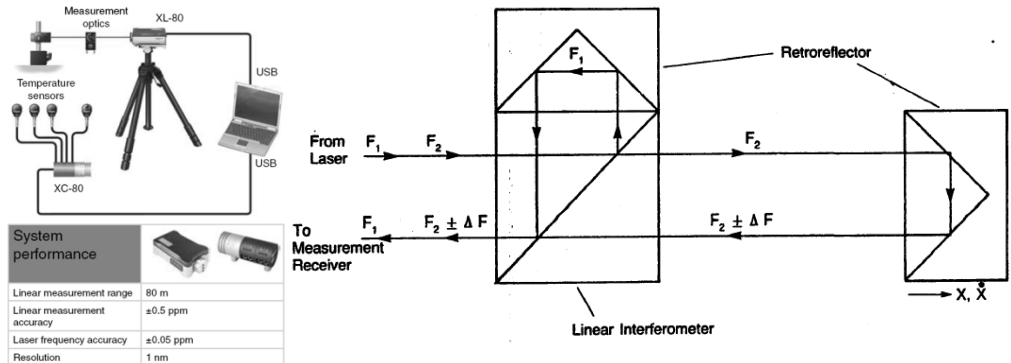

**Figure 7. Renishaw XL-80 laser interferometer system, and simplified optical diagram of interferometric measurement of linear displacement [4].**

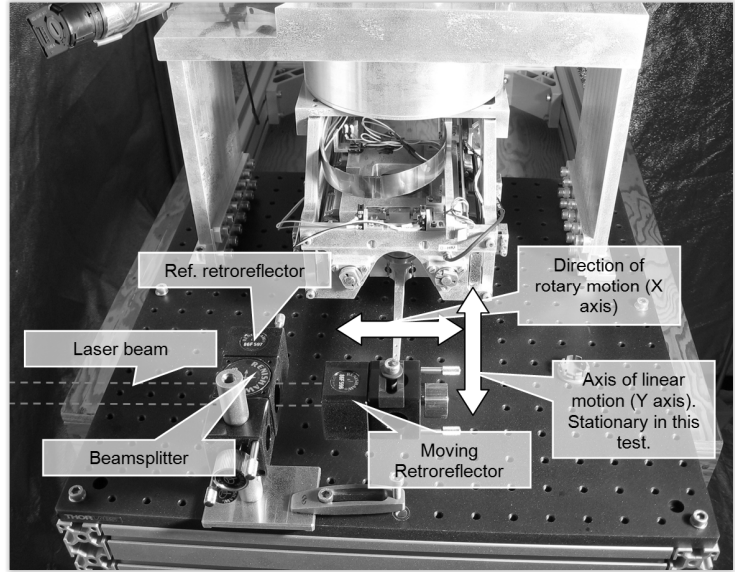

**Figure 8. Rotator cold test, linear interferometric measurement**

## 2.4 SUMMARY OF TEST RESULTS

As an example of one of the tests, Figure 9 depicts a diagram of the linear actuator error over the full range of travel: 0 to 130 mm. This test was done at a room temperature. The uni-directional repeatability measured in this test was ± 3.3 µm.

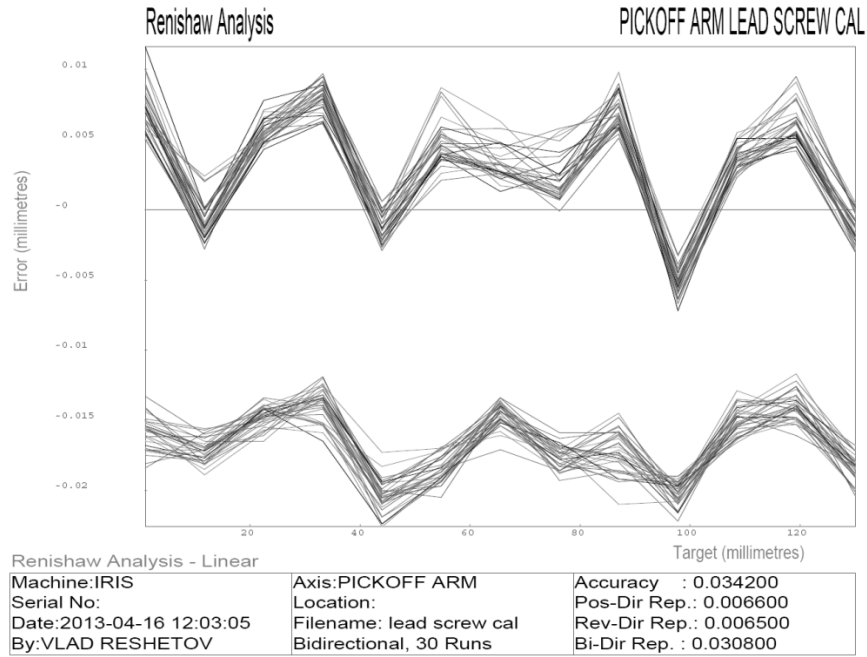

**Figure 9. Room temperature interferometric measurement of the linear actuator travel error over the full range of 0-130 mm, 10 mm steps, 30 bidirectional runs (780 measurements total)**

The summary of the results of the tests are presented in Table 1 and Table 2. All the measured values fall within their respective requirements (detailed in both tables).

**Table 1. Summary of room temperature test results, all values except tilts are ±3 sigma.**

|  | Measured value, [μm] | Measured value, [mas on sky] | Requirement, [μm, μrad] |
|---|---|---|---|
| Repeatability, probe arm rotary actuator | ±3.0 | ±1.36 | ±9.3 |
| Repeatability, probe arm linear actuator | ±3.3 | ±1.65 | ±9.3 |
| Repeatability, collimator | ±3.5 |  | ±10 |
| Maximum tilt around horizontal axis, collimator | 177.6 |  | 500 |
| Maximum tilt around vertical axis, collimator | 124.8 |  | 500 |

**Table 2. Summary of operating temperature (-30ºC) test results, all values are ±3 sigma.**

|  | Measured value, [μm] | Measured value, [mas on sky] | Requirement, [μm] |
|---|---|---|---|
| Repeatability, probe arm rotary actuator | ±1.6 | ±0.73 | ±9.3 |
| Repeatability, probe arm linear actuator | ±2.8 | ±1.27 | ±9.3 |
| Repeatability, collimator | ±4.0 |  | ±10 |

As it is evident from the Figure 9, the linear actuator system exhibits periodic errors. These errors are a combination of the periodic variations in the gearbox ratio, and the periodic variations in the lead screw pitch. The XL-80 system software conveniently allows generation of an error table. Such an error table was compiled and fed into the controller script, and the test was repeated with error correction. Measurement results of the error-corrected actuator are shown in Figure 10. As shown, the use of the error correction table eliminated the effects of the backlash and dramatically increased the accuracy of the system.

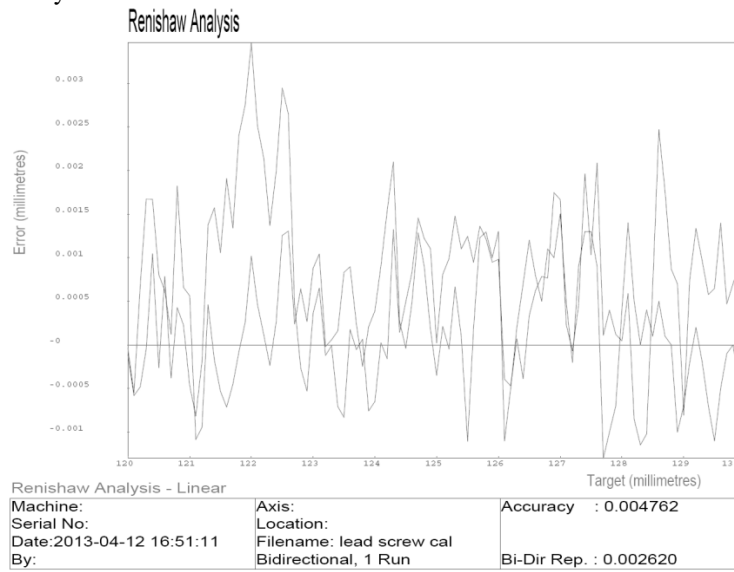

**Figure 10. Error plot on 10 mm travel of the lead screw, forward-reverse cycle. controller script was using an error correction table**

## 3. HANDLING CART

Prior to installation, IRIS will be moved onto the Nasmyth platform and placed on a handling cart to be moved under NFIRAOS. Design of the handling cart was part of our preliminary design stage efforts. The handling cart will be used not only for the installation and removal of IRIS, but also for temporary storage of the 6 ton instrument. In storage position the cart will be reinforced with two additional braces attached to the floor (Figure 11). The braces are necessary to increase stiffness of the cart and therefore to ensure earthquake survival.

The handling cart is designed to run on rails that are fixed to the Nasmyth platform and used to guide the handling cart and IRIS to the correct location underneath NFIRAOS. A lifting mechanism integrated into the cart will allow personnel to raise or lower the instrument into or out of the NFIRAOS port.

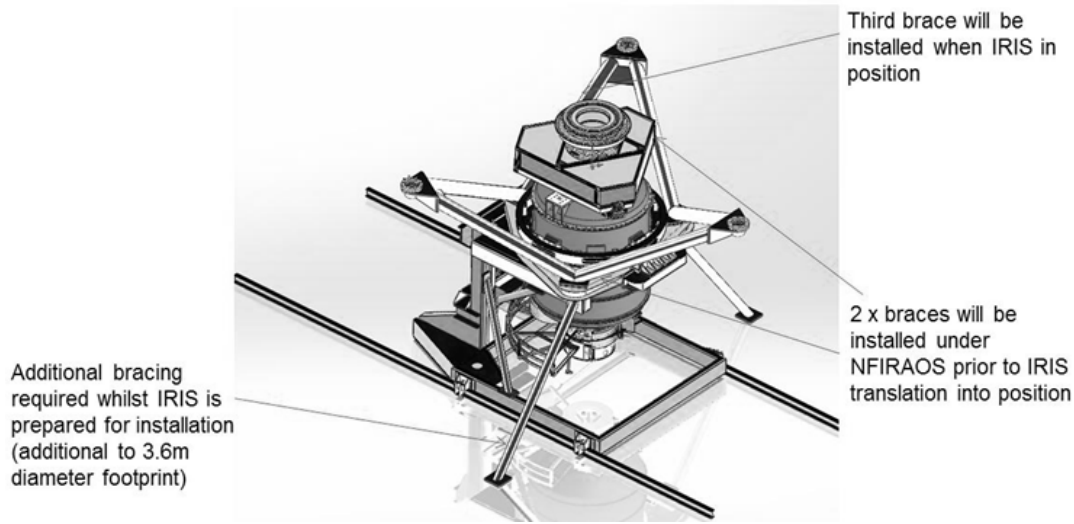

**Figure 11. IRIS on handling cart with storage bracing in place**

## 4. OPTICAL DESIGN

The optical design has remained largely stable since 2010 with some minor modification with respect to changes in the desired wavelength range and the ADC glass selection. The optical design for the OIWFS probe arms is shown in Figure 12 with the TTF mode shown on the left hand side and TT mode on the right hand side. The TTF mode operates with a 2x2 lenslet array inserted at the exit pupil of the collimator lens. In TT mode the lenslet array is removed and an imager lens is inserted into the beam. The magnification of the system in TT mode is required to be twice the magnification in TTF mode. This requires the focal length of the TT imager lens to be twice the focal length of the lenslet array elements.

The TTF lenslet is required to be located at a pupil. The Theta-R opt mechanical configuration of the system requires the exit pupil of the collimator to be located at a back distance from the collimator that allows for
- the 'R' range of the probe (as compensated by the 'Trombone' mirrors),
- the 'Theta' fold mirrors and the space for the ADC,
- and the spacing as required for the imager and lenslet arrays positions.

The converter lens, the first element of the OIWFS, is a negative achromatic doublet used to bring the NFIRAOS very distant exit pupil closer, thus moving the exit pupil after the collimator further as needed with a more compact layout. This converter lens is followed by a fold mirror, and then a field stop (necessary to prevent field overlap in the TTF case). The next element is the collimator that is movable to compensate for NFIRAOS curved focal surface as the OIWFS patrols the field. Two 'trombone mirrors' and two more fold mirrors compensate for R-motion of the probe and direct the beam through the Theta bearing. The mirrors are followed by a Risley prism ADC to compensate for atmospheric dispersion. The imager lens or lenslet arrays are then inserted in the beam to form the image on the detector.

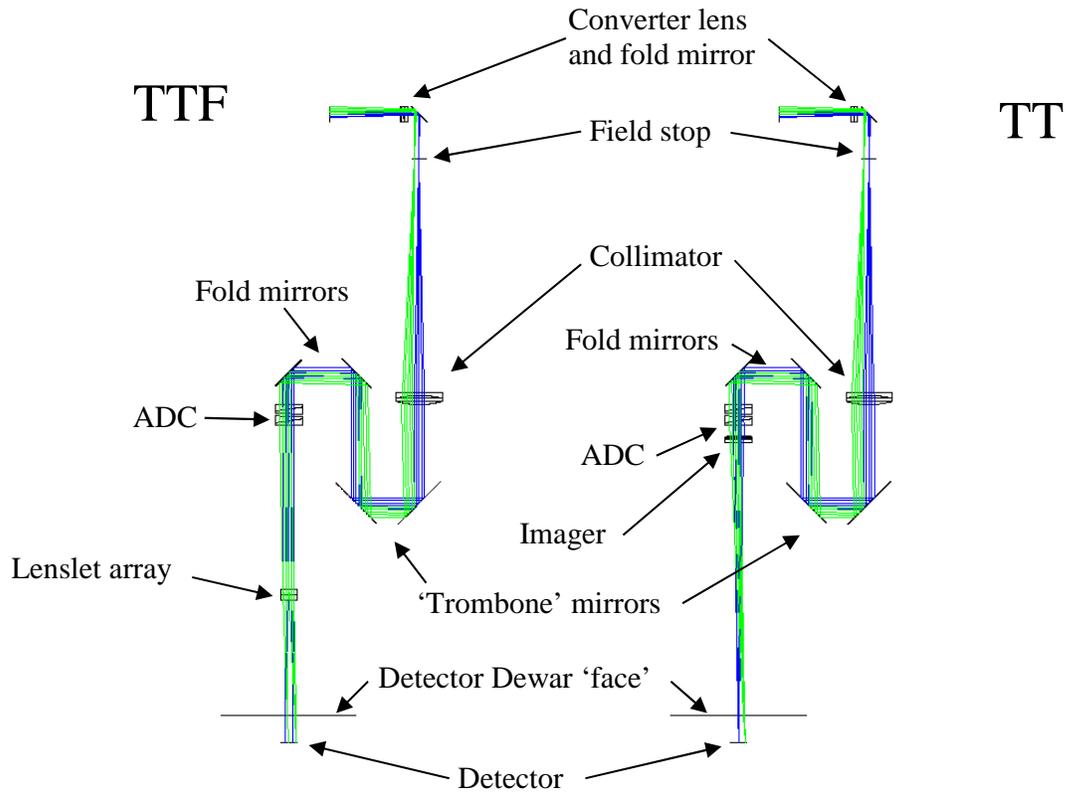

**Figure 12 Annotated layouts of the TTF and TT configurations of the IRIS OIWFSs**

The optical design has been updated for a wider wavelength range, from a design for the J+H bands to a design covering the J+H+Kshort band (1.16um to 2.31um). The optics were optimized for the wider bandwidth without a loss in image quality. The ADC glass selection was re-optimized for the wider band pass, and the optimal glass pair over this range was found to be S-NPH2 with S-BAL42. The residual color of the OIWFS ADC is shown in Figure 13. An interesting consequence of this wider band pass was this new glass combination had a lower residual color than the S-NPH2 with SPINEL.

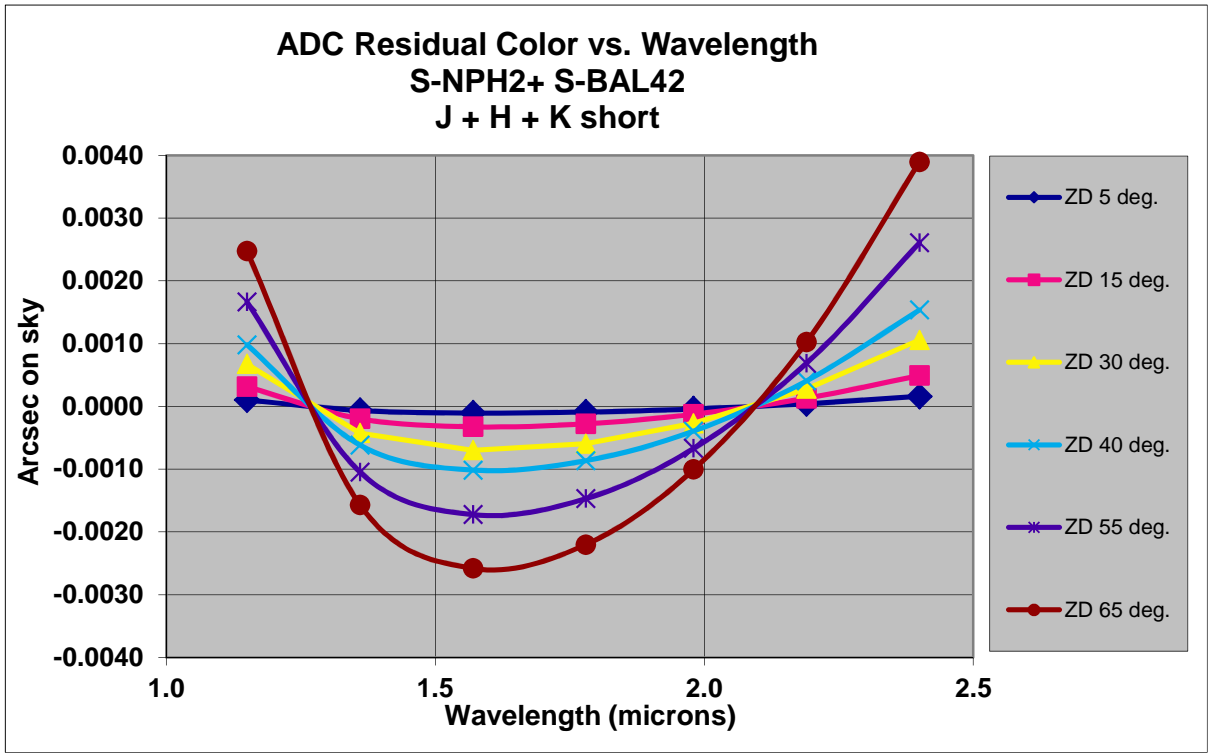

**Figure 13: OIWFS residual color over J+H+Kshort**

## 5.  SPINEL ADC GLASS STUDY

One of the risks identified during the conceptual design phase was the design and performance of the atmospheric dispersion corrector (ADC). There is one ADC located in each probe arm of the OIWFS, plus two more in IRIS. Drew Phillips has investigated many glass combinations for ADCs in IRIS [3]. One of the best combinations for the IRIS imager is Ohara's S-NPH2 and Spinel, therefore a study was initiated to see if it was a viable material to be used for all ADC's in IRIS.  The residual dispersion at a zenith angle of 65 degrees is shown for this configuration in Figure 14 over the whole bandwidth (top) and over J, H, and K-bands individually (bottom).

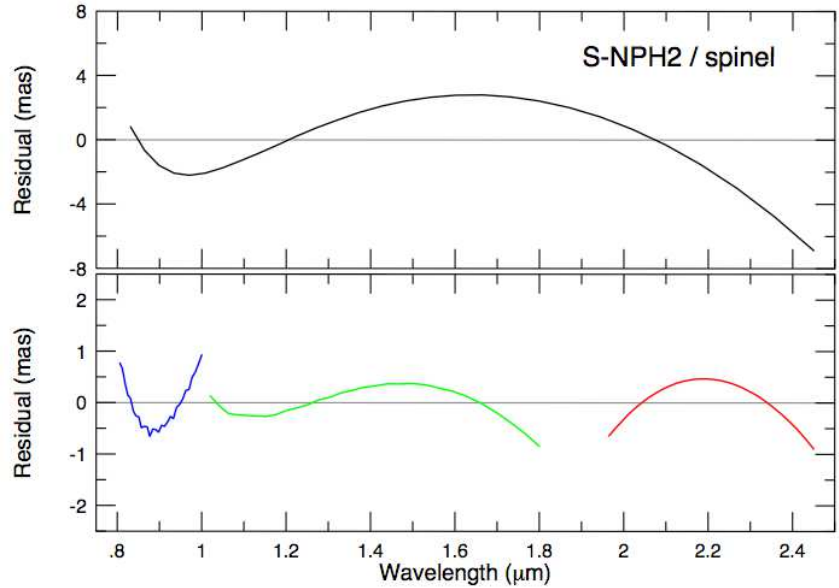

**Figure 14**: Residual dispersion (mas) of the glass combination S-NPH2 and spinel: (top) optimized for the full bandwidth from 0.8-2.4 um, and (bottom) optimized over individual bands J, H, K

Spinel ($MgAl_2O_4$) is a transparent polycrystalline ceramic. It has excellent transmission properties from the UV to beyond 5.5 um. Spinel is currently being produced in commercial form by sintering, hot pressing and hot isostatic pressing (HIP) operations [4]. Optical grade spinel has only been available for a few years, so some of the optical properties are unknown, unreliably measured, or hard to reproduce consistently during fabrication. For these reasons, we began an investigation of the material.

There were three performance measurements we planned to make. First, measure the homogeneity of the material. Poor index homogeneity results in increased wavefront error. Second, measure the refractive index. The manufacturing process can result in different indices batch to bath. Third, quantify the scattering properties. Forward scatter by the crystalline structure was raised early as a possible show-stopper.

We acquired several polished material samples from two vendors. The first were 25 mm diameter by ~14 mm thick windows fabricated by a venture capital R&D company that developed the process of making spinel, and they are limited in the sizes they can produce. Our blanks were polished with the following guidelines: polish 1/10 wave flatness, wedge < 1 arcmin, maximize thickness.

The second sample came from a company that has been working to commercialize the process. Vendor 2 has data on the sample, and felt their process is mature enough for consistent results. The sample we received was 100 mm diameter by ~ 12 mm thick and already polished.

### 5.1 Visual inspection

Visual inspection of the first samples revealed noticeable defects in the surface and bulk material. Defects appeared both light and dark and of ~ 1 mm or smaller in size. A microscope image (Figure 2) shows a number of dark defects on the surface and within the substrate. Some of the imperfections are dark, opaque spots, while others (surface defects) appear to be large voids, or chips, where the glass may have been removed from the substrate during polishing.

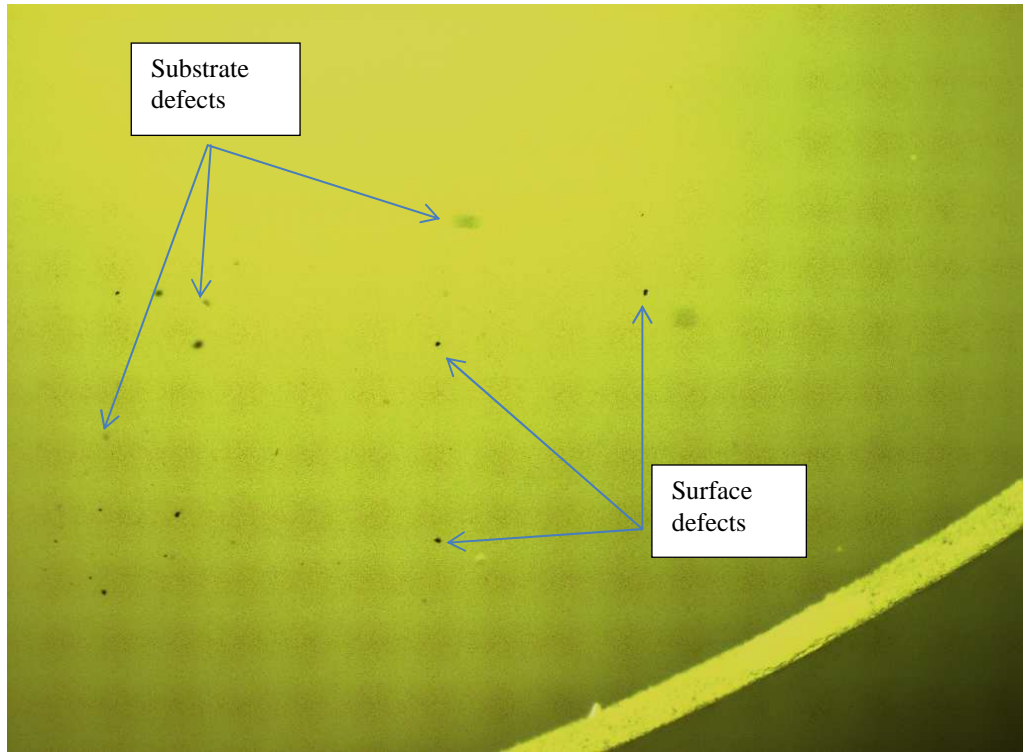

**Figure 15: Microscope image at 6x showing the surface (in focus) and substrate defects seen on a spinel sample from Vendor 1.**

Harder to see, and impossible to photograph, is a "ripple" of the surface, possibly due to insufficient polishing, and possibly due to the grain size of the material. The appearance of the surface could also be described as "scaly" or "pithy" like an orange peel. There is also a "sparkle" to the material in direct sunlight. This sparkle is assumed to be due to inclusions within the material.

The sample from Vendor 2, visually, appears to be of higher quality. There is still a fine "ripple" on the surface, but the features seem smaller and are harder to see. Within the material, there is a continuous distribution of visible particulates that creates a haze. There are no dark particulates and no sign of chips on the surface, as with the Vendor 1 sample. The "sparkle" is also significantly reduced.

### 5.2 Homogeneity tests

The homogeneity tests in our lab were performed with a Wyko interferometer running Intelliwave software. The software has a built-in four-stage subtraction feature to measure homogeneity. The four measurement steps are 1) cavity with transmission through spinel, 2) reflection off first surface, 3) internal reflection off second surface, and 4) empty cavity (see Figure 16). An interferogram of the homogeneity is shown in Figure 17, with an optical path difference (OPD) of 33 nm RMS over a 25 mm aperture. The axes in the plot do not represent the diameter of the sample.

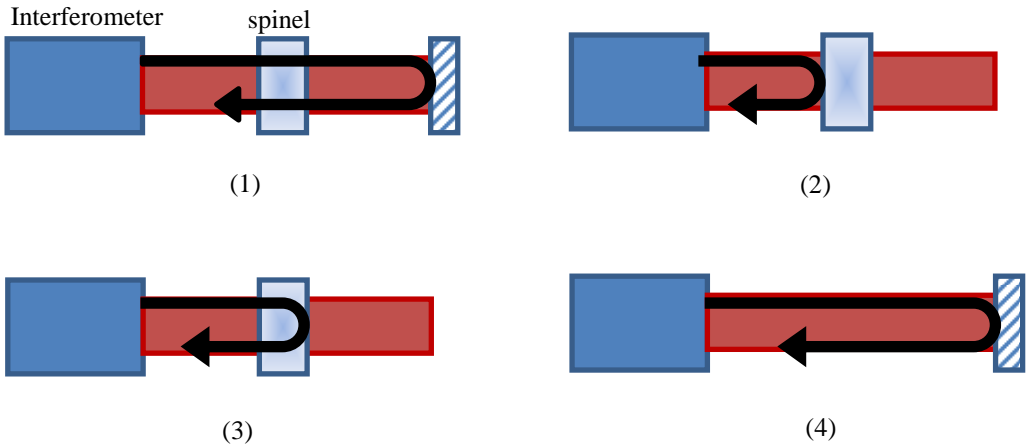

**Figure 16: Graphical depiction of the four measurements taken to determine the homogeneity of the spinel sample: (1) spinel in transmission, (2) reflection off first surface, (3) internal reflection off second surface, and (4) empty cavity.**

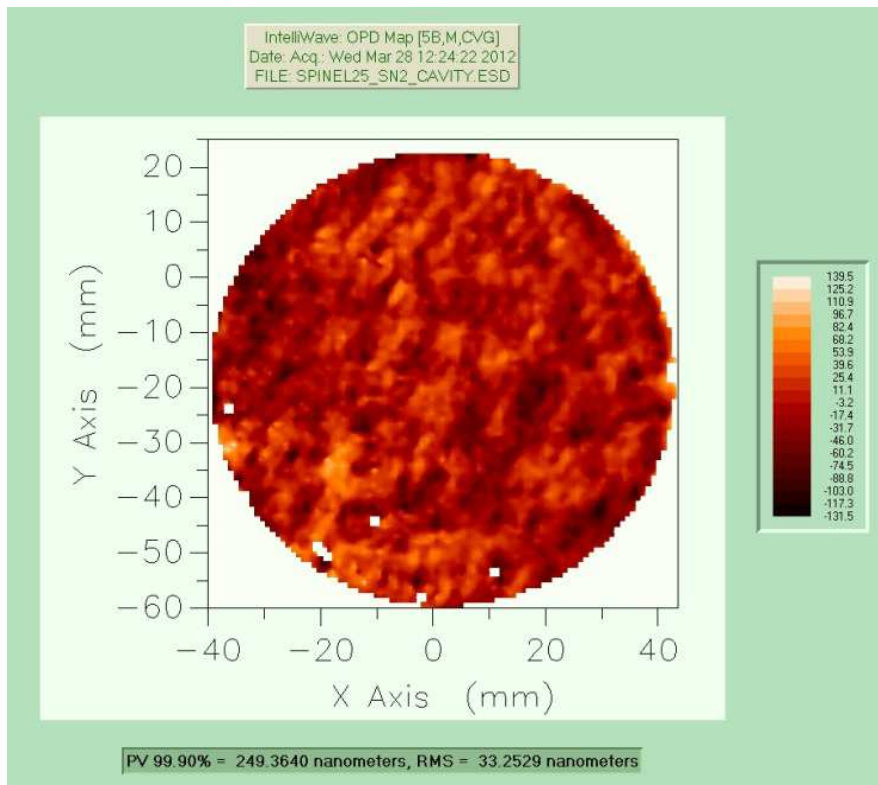

**Figure 17: Homogeneity measurement of one 25 mm sample from Vendor 1 with 33 nm RMS OPD.**

The measurements for our sample from Vendor 2 were done in the same manner. However, the two surfaces of the window were too parallel to get a first or second surface reflection without interference from the other surface. This internal fringing significantly reduced our ability to measure the homogeneity. Our best results gave 55 nm RMS over 100 mm clear aperture, but even these are not reliable. Vendor 2 measured the homogeneity of the parent slab to be 150 nm RMS (piston, tip, tilt, focus removed) over a 330 mm clear aperture, using a "plate and oil" test. With astigmatism also removed, their results were 25-28 nm RMS.

### 5.3 Refractive index

Measuring the refractive index was not pursued, as the homogeneity data did not look promising. Commercial vendors published a single value for the refractive index, but it would be our goal to have multi-wavelength data and multi-temperature data.

### 5.4 Scattering

We did not quantify the scatter in the bulk material, but it is visually significant as can be seen in the Vendor 2 sample in Figure 18. This is an image during alignment of the interferometer of the return images off the transmission flat (center), and the first (left) and second (right) surfaces of the window. The scatter from the second surface, after traversing through the substrate is quite significant.

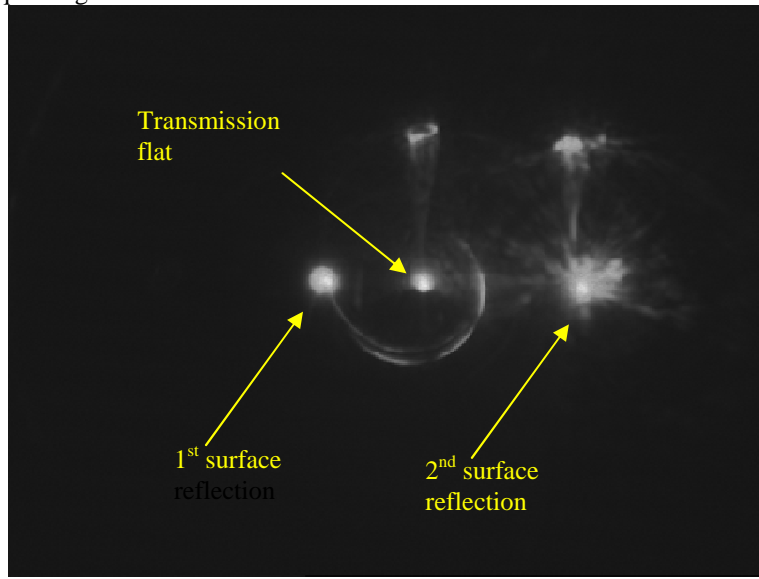

**Figure 18: An image from the interferometer in align mode, showing the bulk scatter of the spinel window from Vendor 2. The center image is from the transmission flat; the left is from the first surface reflection of the spinel window, and the right image is from the second surface of the spinel.**

## 6. SUMMARY

The probe arm prototype served to reduce the risk on being able to patrol the field of view with an appropriate degree of precision that is within the requirements. In terms of the mechanical design future work will include prototyping of the IRIS-NFIRAOS thermal interface, detailed design of the cable wrap, and redesign of the OIWFS probe arms to meet requirements of changed optical prescription.

Our study of spinel showed that the technology is not yet mature enough for a precision application and therefore will not be used for IRIS. Also, with the wider wavelength range for the OIWFS, Spinel is no longer the optimal glass as it does not perform as well over more than one band. The baseline glasses for the ADCs will be S-NPH1/BAL42. However, we will continue to monitor the progress being made on optical grade spinel.

## 7. ACKNOWLEDGEMENTS

The TMT Project gratefully acknowledges the support of the TMT collaborating institutions. They are the Association of Canadian Universities for Research in Astronomy (ACURA), the California Institute of Technology, the University of California, the National Astronomical Observatory of Japan, the National Astronomical Observatories of China and their consortium partners, and the Department of Science and Technology of India and their supported institutes. This work was supported as well by the Gordon and Betty Moore Foundation, the Canada Foundation for Innovation, the Ontario Ministry of Research and Innovation, the National Research Council of Canada, the Natural Sciences and Engineering Research Council of Canada, the British Columbia Knowledge Development Fund, the Association of Universities for Research in Astronomy (AURA), the U.S. National Science Foundation and the National Institutes of Natural Sciences of Japan.## 8. REFERENCES

[1] Loop, D. Reshetov, V., Fletcher, M., Wooff, R., Dunn, J., Moore, A., Smith, R., Hale, D., Dekany, R, Wang, L., Ellerbroek, B., Simard, L., Crampton, D., "The Infrared Imaging Spectrograph (IRIS) for TMT: on-instrument wavefront sensors (OIWFS) and NFIRAOS interface," Proc. SPIE 7735, (2010).

[2] Moore, A., Larkin, J., Wright, S., Bauman, B., Dunn, J., Ellerbroek, B., Phillips, A., Simard, L., Suzuki, R., Zhang, K., Aliado, T., Brims, G., Canfield, J., Chen, S., Dekany, R., Delacroix, A., Do, T., Herriot, G., Ikenoue, B., Johnson, C., Meyer, E., Obuchi, Y., Pazder, J., Reshetov, V., Riddle, R., Saito, S., Smith, R., Sohn, J.M., Uraguchi, F., Usuda, T., Wang, E., Weiss, J., Wooff, R., "The Infrared Imaging Spectrograph (IRIS) for TMT: Instrument Overview," Proc. SPIE 9145, in Ground-based and Airborne Instrumentation for Astronomy V, (2014).

[3] Phillips, A. C., Bauman, B. J. , Larkin, J. E. , Moore, A. M., Niehaus, C. N., Crampton, D., Simard, L., "The infrared imaging spectrograph (IRIS) for TMT; the atmospheric dispersion corrector," Proc. SPIE 7735, (2010).

[4] "Transparent ceramics," http://en.wikipedia.org/wiki/Transparent_ceramics

[5] Donmez, Alkan, "A General Methodology for Machine Tool Accuracy Enhancement Theory, Application and Implementation," Thesis (Ph. D.), Purdue University, 1985.